# Pressure-induced α → ω transition in titanium metal: A systematic study of the effects of uniaxial stress

Daniel Errandonea[1, 2, *], Y. Meng[2], M. Somayazulu[2], D. Häusermann[2]

[1]*Departamento de Física Aplicada-ICMUV, Universitat de València, Edificio de Investigación, c/Dr. Moliner 50, 46100 Burjassot (Valencia), Spain*

[2]*HPCAT, Carnegie Institution of Washington, Advanced Photon Source, Building 434E, Argonne National Laboratory, 9700 South Cass Ave., Argonne, IL 60439, U.S.A.*

We investigated the effects of uniaxial stress on the pressure-induced α → ω transition in pure titanium (Ti) by means of angle dispersive x-ray diffraction in a diamond-anvil cell. Experiments under four different pressure environments reveal that: (1) the onset of the transition depends on the pressure medium used, going from 4.9 GPa (no pressure medium) to 10.5 GPa (argon pressure medium); (2) the α and ω phases coexist over a rather large pressure range, which depends on the pressure medium employed; (3) the hysteresis and quenchability of the ω phase is affected by differences in the sample pressure environment; and (4) a short term laser-heating of Ti lowers the α → ω transition pressure. Possible transition mechanisms are discussed in the light of the present results, which clearly demonstrated the influence of uniaxial stress in the α → ω transition.

---

[*] corresponding author, email: daniel.errandonea@uv.es, Te.: (34) 96 3543432, Fax: (34) 96 3543146

## I. Introduction

Martensitic transformations are abundant in the nature [1 – 4] and have tremendous scientific and technological interest [5, 6]. In particular, the pressure-induced martensitic α (hcp) → ω (hexagonal) transformation in pure titanium (Ti) [7] has significant implications in the aerospace industry because the ω phase formation affects the toughness and ductility of Ti. The α phase (space group number: 194, $P6_3/mmc$) [8] has two atoms per unit cell at (1/3,2/3,1/2) and (2/3,1/3, 3/2) and a *c/a* ratio of ~ 1.58. On the other hand, the ω phase (space group number: 191, P6/mmm) has an interesting crystal structure with three atoms per unit cell at (0,0,0), (1/3,2/3,1/2), and (2/3,1/3,1/2) and a *c/a* ratio of ~ 0.61. Thus, the symmetry of the ω phase is high, 24 point-group operations, the same as for the simple hexagonal structure. The ω phase has a quite open structure and the packing ratio (~0.57) is substantially lower than for the α phase (~0.74). The unusual occurrence of such an open structure under compression has been attributed to a pressure-induced *sp* → *d* electron transfer [9]. The ω phase was observed to be stable up to 40 GPa – 128 GPa [10 –12]. There are however discrepancies on the characterization of the high-pressure post-ω phase.

The occurrence of the pressure-driven α → ω transformation was first observed by Jamieson [13] and has since been studied extensively [10 – 12, 14 - 21]. Room temperature (RT) high-pressure studies of the α → ω transition show a large hysteresis, with the high-pressure ω phase being retained after pressure is released [11, 21]. The onset of the transition has been observed over a wide range of pressures from 2.9 GPa [14] to 11 GPa [17, 19, 20] (see Table I). One of the factors that could be responsible for this scatter in the observed transition pressure ($P_{\alpha \to \omega}$) is possibly due to the variation in the non-hydrostatic conditions in different experiments. In fact, previously it was suggested that shear stress may reduce the transition pressure [22]. However, the combined results from different experiments are inconsistent with this fact (see the previous studies listed in Table I, the highly non-hydrostatic studies of Refs. [10] and [11] reported a higher $P_{\alpha\rightarrow\omega}$ than the quasy-hydrostatic studies of Ref. [21].). Then the question is whether the shear stress explanation is correct and whether other factors play a role in the transition. To answer this question, we conducted a series of experiments in a diamond-anvil cell (DAC) using different pressure media.

The DAC is basically an uniaxial stress device and truly hydrostatic conditions are only obtained when the sample is contained within a fluid pressure medium [23]. At RT a completely hydrostatic environment cannot be sustained above 13 GPa [24, 25] due to solidification of all known pressure medium including helium. Because of this, it is quite possible that the uniaxial stress component [26] of the stress tensor may be quite substantial and different (even at the low pressure where the $\alpha \rightarrow \omega$ transformation was observed in Ti) in the quasy-hydrostatic environment generated in the DAC by many authors [10, 11, 14, 21]. In this study, we examined the effects of uniaxial stresses on the $\alpha \rightarrow \omega$ transition of Ti using synchrotron x-ray powder diffraction. Experiments were performed using a DAC and four different pressure media, which provided different hydrostaticity conditions. We clearly demonstrated that the presence of uniaxial stresses has a significant effect on the structural stability of Ti. We also observed that short term laser-heating of Ti lowers $P_{\alpha\rightarrow\omega}$.

**II. Experimental Details**

The structural stability of Ti under compression was studied up to 16 GPa by angle dispersive powder x-ray diffraction (ADXD). In order to analyze systematically the effects of uniaxial stresses on the α ® ω transition of Ti, we performed four different sets of experiments using a symmetric DAC with the sample loaded under four different pressure transmitting media (argon, 4:1 methanol-ethanol mixture, NaCl, and without pressure medium), which provided different hydrostatiticy conditions [27]. Ti samples, compressed from commercial powder (Alfa Aesar) of stated purity 99.9 %, with a diameter of 50 μm and a thickness of approximately 5 μm were loaded in stainless steel (grade 301) gaskets to perform the studies reported here. The dimensions of the pressure chamber were 100 μm in diameter and 30 μm in thickness. The dominant impurities in the starting samples were O and Fe, with atomic concentrations of 300 and 250 ppm, respectively. Impurity levels from other elements (e.g. Cl, H, N, Mg, and C) were generally much below 100 ppm. It is important to mention that $P_{\alpha \rightarrow \omega}$ has been found to be very sensitive to the oxygen content of the Ti samples [14] when it exceeds 3000 ppm. Therefore, we selected the four studied Ti samples with an identical and negligible oxygen content, in order to ensure that possible differences of their behaviors can be only ascribed to the different pressure environment generated in each experiment.

The experiments were performed using a monochromatic synchrotron radiation source (λ = 0.3875 or 0.4246 Å) at the 16-IDB beamline of the HPCAT facility at the Advanced Photon Source. The monochromatic x-ray beam was focused down, using multilayer bimorph mirrors in a Kickpatrick-Baez configuration [28, 29], to 10 μm by 10 μm. Diffraction images were recorded during 30 sec. with a MarCCD detector and were integrated and corrected for distortions using the FIT2D software [30]. The sample-CCD detector distance was either ≈ 203 mm (when λ = 0.3875 Å) or 215 mm (when λ = 0.4246 Å). Indexing, structure solution, and refinements were performed using the POWDERCELL program package (version 2.4) [31]. The ruby fluorescence technique [32] was applied to measure the pressure.

### III. Results and Discussion

Figure 1 shows ADXD patterns of Ti at selected pressures measured form a sample loaded in a 4:1 methanol-ethanol pressure medium. The six diffraction peaks observed in trace (a) corresponds to the diffraction pattern of the α phase of Ti (α-Ti) at 1 GPa. On increasing pressure, at 10.2 GPa (trace (b)) six new Bragg peaks appear, showing the coexistence of the α and ω phases of Ti at this pressure. The α → ω transition is completed at 14.7 GPa as shown in trace (c). After pressure release the observed transition is not reversible (see trace (d)) in agreement with previous results [21]. Figure 2 shows ADXD spectra of Ti at five different pressures collected from a sample loaded in a NaCl pressure medium. The diffraction pattern shown in trace (a) was measured at 2 GPa and contain two phases: NaCl and α-Ti, with seven Bragg peaks associated to α-Ti. In this case, the onset of the α → ω transition is observed at 6.2 GPa (see trace (b)) and the transition is completed at 14.2 GPa (see trace (d)). In addition, in this sample after pressure release a mixture of the α and ω phases is recovered (see trace (e)). The other two samples studied under different pressure media also show differences regarding the pressure for the first observance of the ω phase, the pressure range of coexistence of both phases, and the crystalline structure of the recovered samples after decompression. The sample studied under argon shows similar results to those of the one studied under a 4:1 methanol-ethanol mixture, occurring the onset of the transition at 10.5 GPa. On the other hand, in the sample studied without pressure medium the results are similar to the results obtained under NaCl but the starting pressure of the transition is the lowest (4.9 GPa). Table I summarizes all our results together with previous results, illustrating the wide range of pressures over which the α → ω transition has been observed.

We now turn our attention to the different behaviors that have been observed for Ti in the range 2 - 11 GPa to show they can be ascribed to different pressure conditions. In order to analyze the effect of uniaxial stresses on the α → ω transition we calculated the volume fraction of the ω phase as a function of pressure in the four samples here studied. The results obtained are shown in Figure 3. There it can be see that in those samples studied under less hydrostatic media (no medium or NaCl) the transition starts at lower pressures than in the other samples and the pressure range of the transition is wider. By taking a look to Figs. 1 and 2, it can be seen that in the sample studied under a NaCl pressure medium the diffraction peaks exhibit a larger half-width than in the sample studied in a 4:1 methanol pressure medium. It has been documented that this broadening of the diffraction peaks is due to more pronounced pressure gradients and to uniaxial stresses [33]. This fact support our idea that uniaxial stresses play an important role on the pressure-driven α → ω transition of Ti. It should be also emphasized that α → ω transformation could be induced by shear stresses during machining [7], which provides additional support to our arguments. The metastability of the α and ω phases observed in shock-wave experiments [17] in the range 10.7 – 14.3 GPa is also in good agreement with our results.

Several transition mechanisms have been proposed for the α → ω transformation [34]. Among them, Silcock's [35] and Usikov's [36] pathways have been the most invoked to describe the α → ω transformation. According to Silcock's mechanism, in each α stacking plane, three of six atoms shuffle by 0.74 Å along [$\bar{1}\bar{1}20$]$_\alpha$, while the other three shuffle on the opposite direction [$11\bar{2}0$]$_\alpha$. This shuffle is accompanied by a strain $e_{xx} = 0.05$ along [$1\bar{1}00$]$_\alpha$ and a strain $e_{yy} = 0.05$ along [$11\bar{2}0$]$_\alpha$ to produce a hexagonal ω cell with the correct *c/a* ratio. In contrast with the direct mechanism proposed by Silcock, Usikov proposed a mechanism with two variants (both having the same strains but different shuffles) which involves a metastable intermediate β (bcc) phase (i.e. α → ω is predicted to proceed as α → β → ω). On the other hand, recently Trinkle *et al.* proposed for the α → ω transformation two pathways (related to Usikov's variants) called TAO-1 ("titanium alpha to omega") and TAO-2 [34]. The TAO-1 mechanism is a direct mechanism in which in the α cell four atoms shuffle by 0.63 Å and two atoms by 0.42 Å, combining this shuffle with strains of $e_{xx} = -0.09$, $e_{yy} = 0.12$, and $e_{zz} = 0.02$ to produce a final ω phase from the α phase [36].

The fact that our experiments systematically demonstrated that uniaxial stresses play a important roll in the α → ω transformation suggests that Silcock's mechanism (which involves the smallest strains) is not appropriate to describe this transformation. In addition, this mechanism involves considerable reconstruction of the lattice which is also in contraction with the martensitic nature that we and previous authors [14] observed for the α → ω transformation. On the other hand, in our studies we did not find any evidence of the existence of a metastable β phase during the α → ω transformation, ruling out Usikov's mechanism. Therefore, our measurements give support to the TAO-1 pathway, as the most likely transition mechanism, in good agreement with recent energy barrier calculations [34].

Another interesting phenomena to comment is the fact than when shear forces are important (no medium or NaCl medium) ω → α transition is observed after some hysteresis under decompression. However, the same fact is not observed in those samples studied under nearly hydrostatic conditions (4:1 methanol-ethanol medium or argon medium), wherein the ω phase is recovered after complete pressure release. According to Sikka *et al.* [7] retention of the high-pressure ω phase is only possible if the uniaxial stress component of the stress tensor is considerably smaller than the transition pressure. This is qualitatively in agreement with the fact that in our case, those samples with narrower diffraction peaks (i.e. smaller uniaxial stresses) do not transform back to the α phase under decompression.

Based upon our results, it is likely to expect that shear stresses present in DAC experiments will also affect the same transition in zirconium and hafnium thereby explaining the observed scatter of the transition pressures reported [9]. In addition, it can be also expected that uniaxial stresses will influence the phase transitions observed at very high pressures in Ti, being the cause of the contradictory results reported by different authors [10 – 12]. Experiments reporting the β, δ, and γ phases of Ti were performed without pressure medium and therefore metastable phases could be formed due to shear stresses as suggested by FP-LAPW calculations [38].

Background corrected x-ray diffraction patterns for all the experiments performed could be reasonably well fitted ($R_{wp} < 0.04$) with the POWDERCELL program [31] considering preferred orientation effects. By fitting all the measured patterns we obtained the pressure dependence for the lattice parameters for both faces, α and ω, which is illustrated in Figure 4. Within the experimental errors, there is no observable effect of pressure medium on the measured unit cell parameters. The data reported in Fig. 4 are in good agreement with previous results [10, 11, 14, 21]. In Fig. 4 it can be also seen that for both phases the lattice compression is anisotropic, with the *a*-axis being clearly more compressible than the *c*-axis. As a consequence of this, the *c/a* ratios of both phases increase with pressure, as shown in Figure 5. For the ω phase, the *c/a* ratio raises from 0.609 at ambient pressure to the ideal ratio, 0.613, at 16 GPa, in good agreement with recent results [10, 11]. The *c/a* ratio of the α phase increases from 1.583 at ambient pressure to 1.622 at 14.5 GPa. All the experimental *c/a* data of the ω phase of Ti can be fitted to the equation:

$c/a = 0.609(1) + 1.1(5) \times 10^{-5} P + 1.3(5) \times 10^{-5} P^2$, where P is in GPa and P < 16 GPa.

And all the experimental *c/a* data of the α phase of Ti can be fitted to the equation:

$c/a = 1.583(1) + 3.2(5) \times 10^{-3} P - 3.5(5) \times 10^{-5} P^2$, where P is in GPa and P < 14.5 GPa.

The molar volume (V) of each phase is plotted as a function of pressure in Figure 6. Using a third-order Birch-Murngahan equation of state [39],

$$P = (3/2) B_0 (x^{7/3} - x^{5/3})[1 + (3/4)(B_0' - 4)(x^{2/3} - 1)], \qquad (3)$$

where $x = V_0/V$, we have determined the RT bulk modulus, its pressure derivative, and the molar volume at ambient conditions of the α phase of Ti, respectively: $B_0 = (117 \pm 9)$ GPa, $B_0' = 3.9 \pm 0.4$, and $V_0 = 10.66 \pm 0.03$ cm$^3$/mol. For the ω phase the following parameters were obtained: $B_0 = (138 \pm 10)$ GPa, $B_0' = 3.8 \pm 0.5$, and $V_0 = 10.48 \pm 0.05$ cm$^3$/mol. These parameters are in good agreement with those previously reported [10, 11, 21]. A volume difference, ΔV/V, of -1.9% is observed between the α and ω phase at ambient pressure, being the same difference of –1.5% at 15 GPa.

Finally, it is interesting to mention that at 5 GPa, two different Ti samples (α phase) loaded under a NaCl pressure medium were double-sided laser heated with the radiation of two Nd.YLF lasers (Photonics GS40, 85 W, TEM$_{01}$ mode, λ = 1053 nm) available at the HPCAT. A detailed description of the HPCAT laser-heating system has been given elsewhere [40]. The aim of these two experiments was the retrieving of the β (bcc) phase of Ti to study its pressure behaviour. In one case, the Ti sample was heated to the stability region of the high-temperature β phase, T = 1750 K, and quenched. In the second case, the Ti sample was heated to a temperature just above the melting [41], T = 2150 K, and quenched. In both cases, samples were laser-heated during approximately one minute, however we could not succeed in quenching the β phase of Ti. In contrast, a mixture of the α and ω phases was obtained at a pressure where only the α phase was observed in RT experiments performed under the same pressure environment. This lowering of the transition pressure after heating suggests that thermal fluctuations induced by the heating could have the same effect as uniaxial stress on the α ® ω transformation of Ti. This phenomenon is consistent with the fact that ω embryos can be stabilized as defects at high temperatures under conditions where the β phase is thermodynamically stable [42]. Upon quenching, the β phase of Ti reverts to the α phase, but the ω embryos could remain stable favoring the onset of the α ® ω transition at lower pressures than in unheated samples.

## IV. Summary

We investigated the effects of the pressure environment generated in a DAC on the α ® ω transformation of Ti. We systematically established that uniaxial stresses lower the transition pressure and observed a coexistence of both phases during a wide pressure range, which depend on the pressure medium used to perform the experiments. We also observed that short term laser-heating of α-Ti favours the α ® ω transformation. We ascribe this fact to the creation of ω embryos during the heating process. The presented results are relevant not only to better the understanding of the transition mechanisms involved in the studied transformation but also to explain the contradictory results reported by different authors at ultra-high pressures [10 – 12].

**Acknowledgments**

This study was made possible through financial support from the Carnegie Institution of Washington and the Spanish government MCYT under grant No. MAT2002-04539-CO2-01. Use of the Advanced Photon Source (APS) was supported by the U. S. Department of Energy, Office of Science, Office of Basic Energy Sciences, under Contract No. W-31-109-Eng-38. Use of the HPCAT facility was supported by DOE-BES, DOE-NNSA, NSF, DOD -TACOM, and the W.M. Keck Foundation. We would like to thank the rest of the staff at the HPCAT at APS for their contribution to the success of the experiments. Daniel Errandonea acknowledges the financial support from the MCYT of Spain and the Universitat of València through the "Ramón y Cajal" program for young scientists. The authors also gratefully thanks F. Jona, S. K. Sikka, and D. Trinkle for making valuable comments to the manuscript.

**References**

[1] P. S. Kotval and R. W. K. Honeycombe, Acta Metall. **16**, 597 (1968).

[2] C. S. Yoo, H. Cynn, P. Söderlind, and V. Iota, Phys. Rev. Lett. **84**, 4132 (2000).

[3] H. Cynn, C. S. Yoo, B. Baer, V. Iota-Herbei, A. K. McMahan, M. Nicol, and S. Carlson, Phys. Rev. Lett. **86**, 4552 (2001).

[4] D. Errandonea, B. Schwager, R. Boehler, and M. Ross, Phys. Rev. B **65**, 214110 (2002).

[5] *Martensitic*, ed. By G. B. Olson and W. S. Owen (ABM, Metals Park OH, 1992).

[6] *Shape Memory Materials*, ed. By K. Otsuka and C. M. Wayner (Cambridge University Press, Cambridge, 1998).

[7] S. K. Sikka, Y. K. Vohra, and R. Chidamberan, Prog. Mater. Sci. **27**, 245 (1982).

[8] K. M. Ho, C. l. Fu, B. N. Harmon, W. Weber, and D. R. Hamann, Phys. Rev. Lett. **49**, 673 (1982).

[9] R. Ahuja, J. M. Will, B. Johansson, and O. Eriksson, Phys. Rev. B **48**, 16269 (1993) and references therein.

[10] Y. Akahama, H. Kawamura, and T. L. Bihan, Phys. Rev. Lett. **87**, 275503 (2001).

[11] Y. K. Vohra and P. T. Spencer Phys. Rev. Lett. **86**, 3068 (2001).

[12] R. Ahuja and L. Dubrovinsky, Poster presented at the 41st EHPRG meeting (Bordeaux, 2003). R. Ahuja and L. Dubrovinsky, "Titanium metal at high-pressure: Synchrotron experiments and *ab initio* calculations", Oral presentation at the SMEC 2003 (Miami. 2003).

[13] J. C. Jamieson, Science **140**, 72 (1963).

[14] Y. K. Vohra, S. K. Sikka, S. N. Vaidya, and R. Chidamberan, J. Phys. Chem. Solids **38**, 1293 (1977).

[15] H. Xia, G. Parthasarathy, H. Luo, Y. K. Vohra, and A. L. Ruoff, Phys. Rev. B **42**, 6736 (1990).

[16] R. F. Trunin, G. V. Simakov, and A. B. Medvedev, High Temp. **37**, 851 (1999).

[17] C. W. Greef, D. R. Trinkle, and R. C. Albers, J. Appl. Phys. **90**, 2221 (2001).

[18] A. Jayaraman, W. Klement, and G. C. Kennedy, Phys. Rev. **131**, 644 (1963).

[19] R. G. McQueen, S. P. Marsh, J. W. Taylor, J. N. Frotz, and W. J. Carter, in *High Velocity Impact Phenomena*, ed. by R. Kinslow (Academic, New York, 1970).

[20] G. T. Gray, C. E. Morris, and A. C. Lawson, *Omega phase formation in titanium and titanium alloys*, in *Titanium '92: Science and Technology*, ed. F. H. Froes and I. L. Caplan, (Warrendale, PA, Minerals Metals & Materials Society) p. 225-232 (1993).

[21] L. C. Ming, M. Manghnani, and M. Katahara, Acta Metall. **29**, 479 (1981).

[22] V. A. Zilbershtein, N. P. Chistotina, A. A. Zharov, S. S. Grishina, and E. I. Estrin, Fiz. Met. Metalloved. **39**, 445 (1975).

[23] A. Jayaraman, Rev. Mod. Phys. **55**, 65 (1983).

[24] T. S. Duffy, G. Shen, D. L. Heinz, J. Shu, Y. Ma, H. K. Mao, R. J. Hemley, and A. K. Singh, Phys. Rev. B **60**, 15063 (1999).

[25] P. Loubeyre, R. Le Toulec, D. Häusermann, M. Hanfland, R. Hemley, H. K. Mao, and L. W. Finger, Nature **383**, 702 (1996).

[26] The stress tensor in the center of a diamond-anvil cell can be written as
$$\boldsymbol{\sigma} = \begin{bmatrix} \sigma_P & 0 & 0 \\ 0 & \sigma_P & 0 \\ 0 & 0 & \sigma_P \end{bmatrix} + \begin{bmatrix} -t/3 & 0 & 0 \\ 0 & -t/3 & 0 \\ 0 & 0 & 2t/3 \end{bmatrix}$$
where $\sigma_P$ is the mean normal stress or pressure and $t$ is the uniaxial stress component.

[27] At the transition pressure, argon provides nearly hydrostatic conditions and the sample studied without pressure medium was put under highly non-hydrostatic

conditions. The hydrostaticity of the pressure medium decreases following the sequence: argon ® 4:1 methanol-ethanol ® NaCl ® no pressure medium.

[28] R. Signorato, *Proceedings of 44th SPIE Anual Meeting*, (Denver, Colorado, 1999) [3773-06].

[29] R. Signorato, T. Ishikawa, and J. Carre, *Proceedings of 47th SPIE Anual Meeting*, (San Diego, California, 2001) [4501-09].

[30] A. P. Hammersley, S. O. Svensson, M. Hanfland, A. N. Fitch, and D. Häusermann. High Pres. Res. **14**, 235 (1996).

[31] W. Kraus and G. Nolze, J. Appl. Crystallogr. **29**, 301 (1996).

[32] H. K. Mao, J. Xu, and P. M. Bell, J. Geophys. Res.: Solid Earth **91**, 4673 (1986).

[33] M. Gauthier, High. Pres. Res. **22**, 779 (2002).

[34] D. R. Trinkle, R. G. Hennig, S. G. Srinivasan, D. M. Hetch, M. D. Jones, H. T. Stokes, R. C. Ahrens, and J. W. Wilkins, Phys. Rev. Lett. **91**, 025701 (2003).

[35] J. M. Silcock, Acta Metall. **6**, 481 (1958).

[36] M. P. Usikov and V. A. Zilbershtein, phys. stat. sol. (a) **19**, 53 (1973).

[37] See Figs. 1 and 3 of Ref. [34] for a nice representation of the TAO-1 mechanism.

[38] K. D. Joshi, G. Jyoti, S. C. Gupta, and S. K. Sikka, Phys. Rev. B **65**, 052106 (2002).

[39] F. Birch, Phys. Rev. **71**, 809 (1947).

[40] D. Errandonea, M. Somayazulu, D. Häusermann, and H. K. Mao, Journal of Physics: Condensed Matter **15**, 7635 (2003).

[41] D. Errandonea, B. Schwager, R. Ditz, C. Gessmann, R. Boehler, and M. Ross Phys. Rev. B **63**, 132104 (2001).

[42] J. M. Sanchez and D. De Fontaine, Acta Metall. **26**, 1083 (1978).

**Figure Captions**

Figure 1: x-ray diffraction patterns of Ti at different pressures: (a) 1 GPa, (b) 10.2 GPa, (c) 14.7 GPa, and (d) 1 GPa after pressure release. The sample was loaded using 4:1 methanol-ethanol as pressure transmitting medium. Miller indices corresponding to the $\alpha$ and $\omega$ phases of Ti are indicated.

Figure 2: x-ray diffraction patterns of Ti: (a) 2 GPa, (b) 6.2 GPa, (c) 13.2 GPa, (d) 14.2 GPa, and (e) 2.7 GPa after pressure release. The sample was loaded using NaCl as pressure transmitting medium. Miller indices corresponding to the $\alpha$ and $\omega$ phases of Ti are indicated. NaCl and gasket (*) diffraction lines are also shown.

Figure 3: Relative amounts of $\omega$ to $\alpha$ Ti at high pressures clearly showing that completion of the $\alpha$-to-$\omega$ transition depend on the sample environment. (?) No pressure medium, (o) NaCl pressure medium, ($\Delta$) 4:1 methanol-ethanol , and ($\nabla$) argon pressure medium. It contains the data obtained during compression (solid symbols) and decompression (empty symbols). Solid lines are just a guide to the eye.

Figure 4: The pressure dependence of the lattice parameters for the $\alpha$ and $\omega$ phases. Different symbols correspond to experiments carried out under different pressure media as in Fig. 3. Solid symbols correspond to compression data and empty symbols to decompression data. Black diamonds were taken from references [11] and [14], gray diamonds from ref. [10], and white diamonds from Ref. [21]. Solid lines are just a guide to the eye.

Figure 5: The pressure dependence for the *c/a* ratios for the α (solid symbols) and ω (empty symbols) phases of Ti. Different symbols correspond to experiment carried out under different pressure media as in Fig. 3. Diamonds were taken from references [11] and [14], pentagons from ref. [10], and hexagons from Ref. [21]. Solid lines are fits to all data as described in the text.

Figure 6: The pressure dependence of the atomic volume from the α (solid symbols) and ω (empty symbols) phases of Ti. Different symbols correspond to experiments carried out under different pressure media as in Fig. 3. Diamonds were taken from references [11] and [14], pentagons from ref. [10], and hexagons from Ref. [21]. Solid lines represent the results of the least-squares fits of the Birch-Murnaghan equation of state to the data of both phases.

**Table I:** Summary of α ® ω transformation pressure data for Ti

| Pressure medium | $P_{\alpha \to \omega}$ [GPa] | Pressure release phase | Experiment |
|---|---|---|---|
| argon | 10.5 ® 14.9 | ω | This work |
| 4:1 methanol-ethanol | 10.2 ® 14.7 | ω | This work |
| NaCl | 6.2 ® 14.2 | α + ω | This work |
| no pressure medium | 4.9 ® 12.4 | α + ω | This work |
| no pressure medium | 2.9 | not given | Ref. [14] |
| no pressure medium | 7.4 | not given | Ref. [10] |
| no pressure medium | 9 | ω | Ref. [11] |
| 4:1 methanol-ethanol | 4 - 7 | ω | Ref. [21] |
| shock-induced | 11 | ω | Ref. [17, 19] |

**Figure 1**

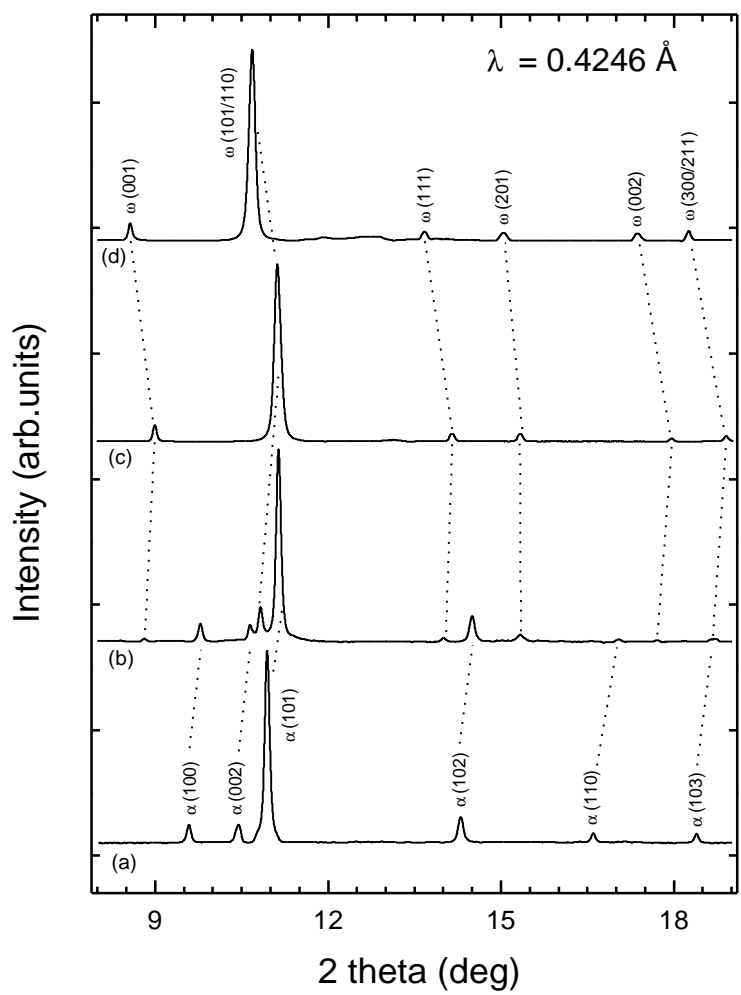

**Figure 2**

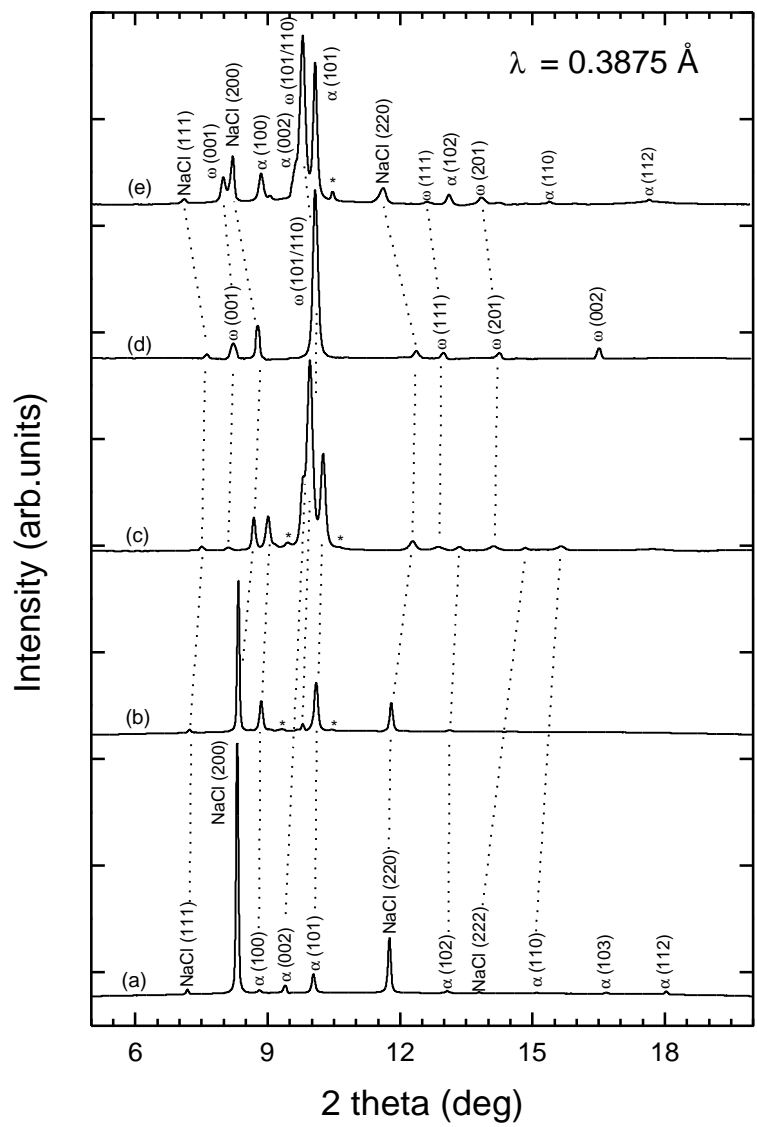

**Figure 3**

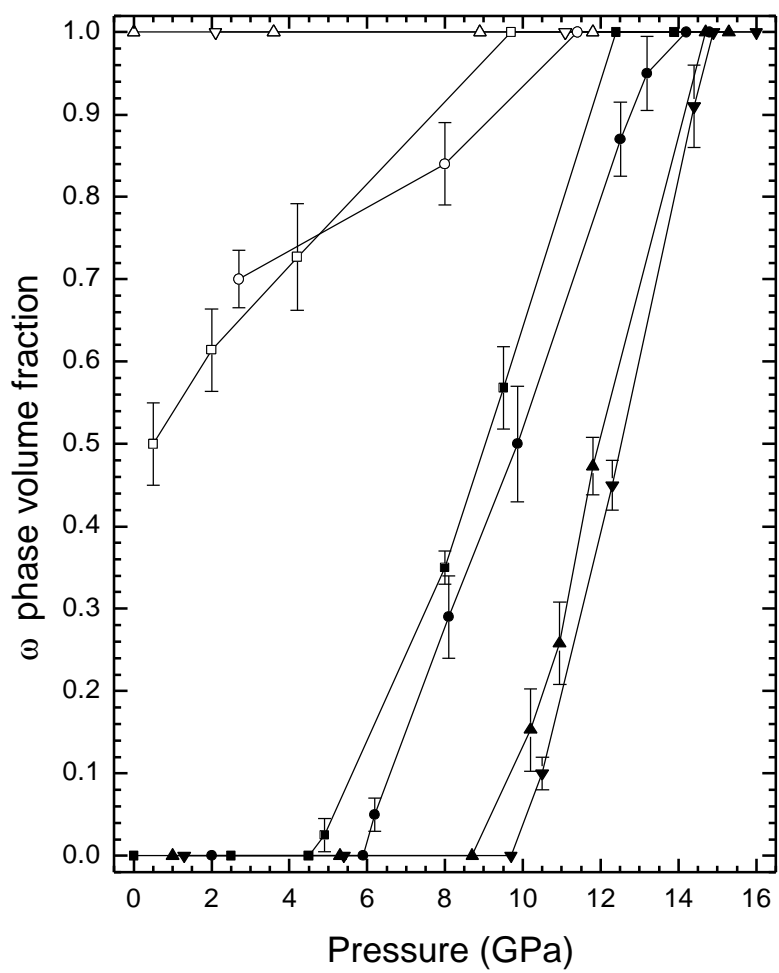

**Figure 4**

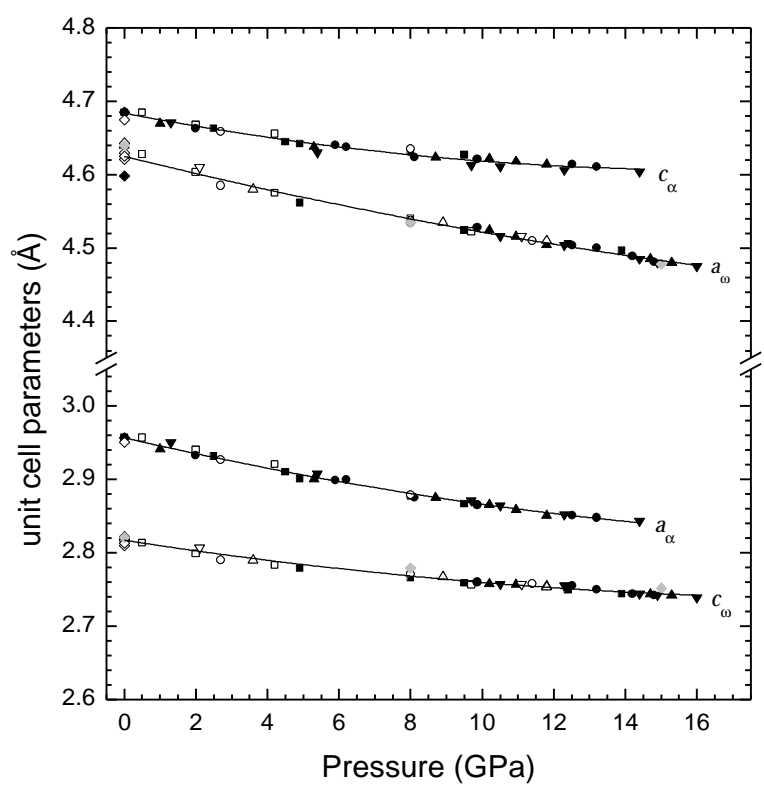

**Figure 5**

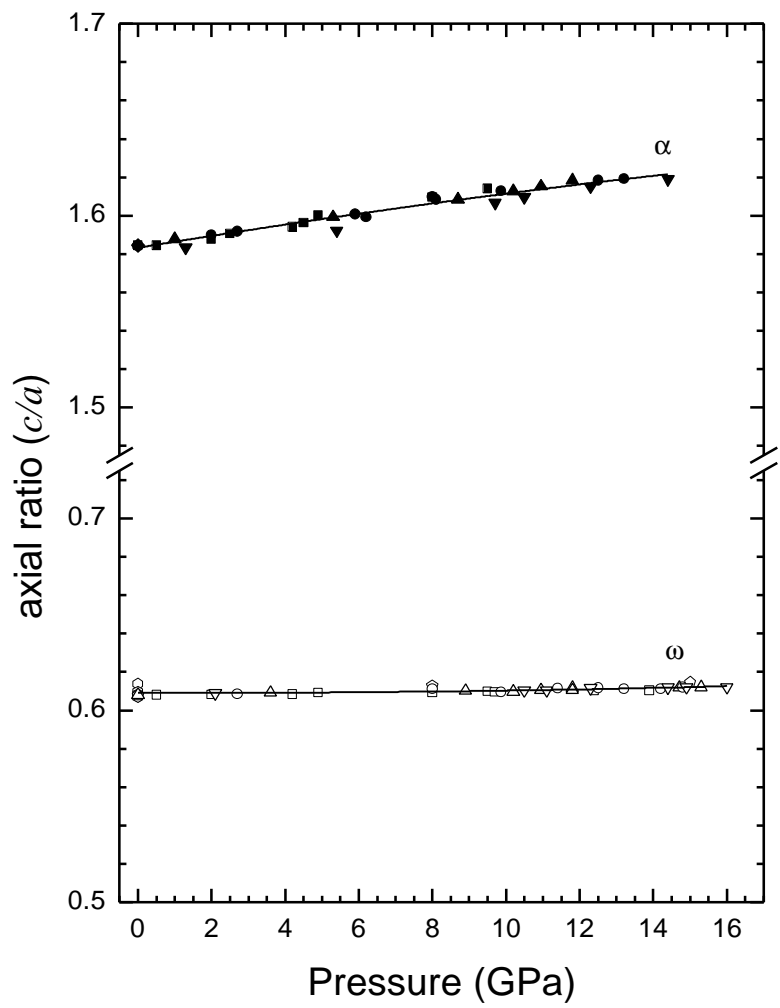

**Figure 6**

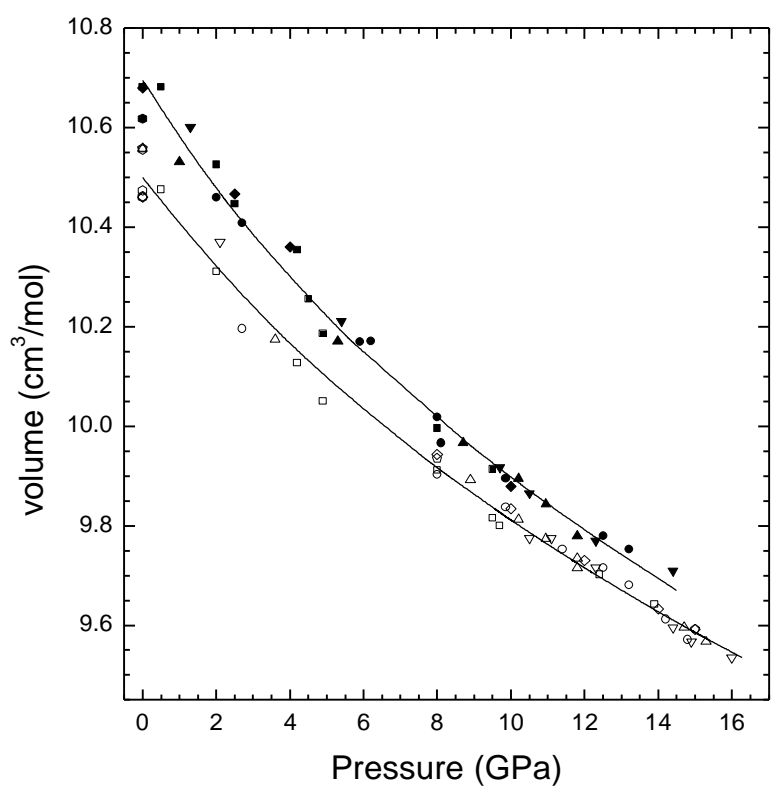